%%%PLAIN TEX FILE OF THE MANUSCRIPT
\magnification=\magstep1
\baselineskip=24 true pt
\hsize=16 true cm
\vsize=22 true cm

\centerline {\bf Resonant Collisions in Four Dimensional Reversible}
\centerline {\bf Maps: A Description of Scenarios.}

\bigskip

\centerline{Avijit Lahiri,}
\centerline{Vidyasagar Evening College, Calcutta 700 009}

\centerline{Ajanta Bhowal,}
\centerline{Saha Institute of Nuclear Physics, Calcutta 700 064}

\centerline{and}

\centerline{Tarun K. Roy,}
\centerline{Saha Institute of Nuclear Physics, Calcutta 700 064}

\bigskip
\noindent {\bf Abstract:} We define a resonant collision of order $k(\geq 1)$ in
 a family of 
four-dimensional reversible maps. For any specified $k$, the bifurcation 
{\it secnario} is the collection of the different possible {\it types} of 
bifurcation of a symmetric fixed point that may be encountered through 
various choices 
of parameters describing the family of maps under consideration. We adopt a 
perturbative approach, coupled with numerical iterations around orbits 
obtained perturbatively, to explore phase space structures in the immediate 
vicinity of a resonant collision and thereby to obtain a description of the 
possible scenarios for different values of $k$. The phase space structures 
typically involve bifurcating periodic orbits, families of invariant curves, 
and tori, and present interesting possibiliries, especially around the 
'secondary bifurcations' of the periodic orbits (see below). Based on the 
results of the perturbative and numerical approach we conjecture that 
three distinct scenarios are involved for the cases $k = 2, 3, 4$ 
respectively, while 
there exists a fourth distinctive scenario common to all $k > 4$, and we 
present what we believe to be a reasonably exhaustive description of these 
scenarios. The case $k = 1$ involves bifurcations of fixed points rather 
than of periodic orbits, and has been investigated numerically in a previous 
paper.

\vfil\eject

\noindent {\bf I. INTRODUCTION.}
\bigskip
Resonances in 4D symplectic and reversible maps occurring through the
collision of a quartet of multipliers on the unit circle and their 
subsequent departure from the circle are, in
a sense, counterparts of corresponding resonances  
in 4D Hamiltonian and reversible flows. 

In the context of Hamiltonian flows a resonance of this type involves 
the collision 
on the imaginary axis of a quartet of eigenvalues associated with a 
fixed point, and their subsequent departure therefrom, and is termed 
the $1:-1$ resonance, where the minus sign indicates that the pairs of 
eigenvalues involved are of opposite Krein signature [1]. An alternate 
nomenclature is 1:1 non-semisimple resonance (as opposed to the 1:1 
semisimple resonance where the eigenvalues do not leave the imaginary axis).
There is a long and distinguished history of investigations on the 
bifurcation associated with this resonance, 
originating in the restricted three-body problem ([2], see [3] for
details and for historical notes), and this bifurcation has come to be 
known as the 
Hamiltonian Hopf bifurcation [3]. It is encountered in the vicinity of the 
Routh's critical mass ratio in the restricted three body problem [4], as 
well as in a spinning `orthogonal' double pendulum [5], and in an atom 
placed in a rotating electric field [6].

Analogous resonance and bifurcation phenomena are encountered in reversible 
flows, occurring in the theory of non-linear oscillations ([7,8]~; see 
references therein). While the Krein theory is not applicable to 
non-Hamiltonian reversible flows, still these flows share many 
characteristic features of Hamiltonian ones close to symmetric fixed 
points and periodic orbits (see [7,8] for details).
                                                               
In this context, it is useful to look into analogous resonances in 4D 
symplectic and reversible $ maps$ involving a quartet of multipliers 
of a fixed point on the unit circle where, generically speaking, results 
resembling those in flows are obtained. However, 4D maps are, truly 
speaking, representatives of dynamics of 6D flows, and this shows up in 
an essential complication in collisions in 4D maps as compared to 4D 
flows~: in case of maps one has to distinguish between the $rational$ 
and $irrational$ collisions.

While a collision itself is a resonance phenomenon, a rational collision 
indicates the presence of an additional resonance, and may be termed a 
`resonant' collision to distinguish it from an irrational or `non-resonant'
collision, the latter being generic in the measure-theoretic sense.

Irrational collisions in 4D reversible maps have been studied in [9,10,11]. 
In the Hamiltonian setting they have been investigated in [12] with 
rigorous results. Rational collisions at $\pm i$ on the unit circle, 
on the other hand, have been analysed in [13] for sympletic maps, 
while a rational collision at $\pm i$ for a certain family of reversible
maps analysed in [14] revealed features quite similar to the symplectic
$\pm i$ collision (see also [15]). Interesting numerical results on 
bifurcations near rational and irrational collisions are to be found in [16].

A collision of a  quartet of multipliers at $e^{\pm i\phi}$ (for a fixed 
point of a symplectic map or a symmetric fixed point of a reversible map), 
with $\phi =2\pi /k$  ($k$ positive integer)
will be termed a resonant collision of order $k$. 

While a collision of multipliers at an irrational angle is generally accompanied
by the bifurcation of families of invariant curves, resonant (or rational) 
collisions are typically characterised by bifurcations of periodic orbits. 

In this paper we present perturbation-theoretic and numerical 
results for resonant collisions of order $k$ ($k=2,3,\cdots ,6$), 
from which we shall see that each of the cases $k=2,3,4$ involves a 
characteristic bifurcation `scenario' (see below) 
while the cases $k>4$ present a common but distinct scenario. In
a sense, the situation resembles the bifurcation of periodic orbits in 2D
symplectic maps where resonances of order $k$ are seen to lead to different 
scenarios for $k$=2,3 and 4 while there occurs a fourth and distinct
scenario for any $k>4$ [4, 18].     

As mentioned above, the bifurcation of period-4 orbits in 4D symplectic 
and reversible maps has been studied in the literature in some detail [13,14]. 
While the dynamics in phase  space around the periodic orbits was not explored 
in these works, preliminary results in this regard were presented in [15] 
where families of invariant curves around the bifurcating
period-4 orbits were investigated. An interesting feature in this context
is the so called `secondary bifurcation' [13,15] whereby elliptic invariant 
sets involved in  a rational angle bifurcation change stability causing a 
further change in local phase space structure. As will be seen in the 
following, this feature occurs commonly in the vicinity of resonant 
collisions of other orders as well.

We plan this paper as follows~:

\noindent In section II we specify the family of maps to be studied and 
explain a few introductory notions. Sections III, IV and V deal with 
resonant collisions of order 2, 3 and 4 respectively, the first and the 
last of these being  
brief partial summaries, with necessary elaborations in a few instances, 
of results of ref. [15]. Section VI includes a few preliminary results 
on resonant collisions of orders 5 and 6. Section VII contains a synopsis and 
concluding remarks.
\bigskip
\noindent {\bf II. RESONANT COLLISIONS IN TWO-PARAMETER FAMILIES OF 4D REVERSIBLE MAPS.}
\bigskip

In this paper we consider situations involving 
two-parameter families of 4D reversible maps $A_{q,\epsilon}$ such that~:

\noindent (i) $A_{q,\epsilon}$ has a symmetric fixed point at the origin
for each $q$, $\epsilon$ in some domain D of the parameter space (which is 
2D, involving parameters $q,\epsilon$)
containing the point $q=0$ and $\epsilon =0$;

\noindent (ii) $\bar A_{0,0}$ has eigenvalues $e^{\pm {2\pi i\over k}}$ 
($k$=integer$ \geq 2$)
with each eigenvalue repeated twice (we call it a resonant collision of 
order $k$) or an eigenvalue $-1$ with multiplicity 4 (we call it a resonant 
collision of order 2).

\noindent [Remark~: For definitions of `reversible map' and `symmetric 
fixed point', see ref.s [9],[10]~; here $\bar A$ denotes the linearisation 
of A at the fixed point at the origin, the relevant parameter values being 
indicated through subscripts.]

\noindent (iii) for any $q$, $\epsilon$ in the domain D, other than (0,0)
the linearisation $\bar A_{q,\epsilon}$ at the origin has distinct 
eigenvalues.

\noindent (iv) the Jordan block structure of $\bar A_{0,0}$ contains either
a real Jordan block of order 4 ($k=2$) or two complex Jordan blocks each of 
order 2 ($k>2$).

\noindent It will be seen from the following that for each family of maps 
of the above type, the parameter space is typically 
divided into several regions by two curves as shown in fig.1 (these parameter 
regions have been marked for future reference as $P1a$, $P1b$, $P2$, 
and $P3$), where the 
parametrisation has been chosen in such a way that the curves are represented 
by the equations $$\epsilon =0,~~~q^2+\epsilon =0 \eqno (1a,b),$$ 
these curves having a tangency at
(0,0). The local structure of the phase space around the fixed point is
distinct in each of these regions. When any one of the boundary curves is
crossed at a point other than (0,0) in the parameter space the change in the 
local phase space corresponds to a codimension-1 bifurcation. The point 
(0,0) in the parameter space, where two boundary curves meet, is 
degenerate, and the associated bifurcation is of codimension 2.

The local phase space structure
in the vicinity of the fixed point for any $A_{q,\epsilon}$ is essentially 
described in terms of stability  type of
the fixed point and the existence and stability type  of the bifurcating 
periodic 
points (a periodic point is said to bifurcate from the fixed point if 
it tends 
to and merges with the fixed point as some particular point in the parameter
space is approached). Additionally, a description of families of invariant 
curves and tori around the
periodic point may have to be included for a reasonably complete description
of the local phase space structure. Depending on the nonlinear terms, each 
family
will be found to involve a certain number of {\it {types}} of changes in the
local phase space structure 
as the
different regions around the point (0,0) in the parameter space are approached 
from one another. The 
possible
different types depend on the order of the resonant collision. Thus we get
different {\it {scenarios}} of  bifurcations, each scenario  involving 
its own types of change in the local phase space structure. As already 
stated, the situations
with $k=2,3$ and $4$ are found to correspond to three distinct scenarios and there is a 
fourth
scenario which is common to the situations involving all resonant collisions
of order $k\geq 5$. We describe below the different scenarios obtained from
our analysis. However, the description is not complete since each scenario
typically involves certain degenerate types which have been left outside the 
purview of our investigations. These degenerate types are described by normal
forms involving higher degree terms compared to the normal forms of the 
nondegenerate cases. While we have not attempted a normal form analysis of 
the resonances in the present work we conjecture that, {\it modulo} the degenerate 
cases, the scenarios indicated above do exhaust the typical behaviours of families
of 4D reversible maps satisfying criteria (i)-(iv) above.

\bigskip
\noindent {\bf III. SECOND ORDER RESONANT COLLISIONS AND }
\noindent {\bf PERIOD-2 ORBITS}
\bigskip

\noindent In order to investigate the second order resonant collision 
i.e., a collision near the point $-1$ on the unit circle, we consider the
two parameter family of maps $A_{q,\epsilon}$, described in the form of the 
following fourth order difference equation~:
$$x_{n+2} + x_{n-2} -2(q-2)(x_{n+1}+x_{n-1}) + (q^2-4q+6+\epsilon)x_n =
\beta x_n^3\eqno(2),$$
\noindent where $\epsilon,~q $ are small bifurcation parameters and
$\beta\neq 0$ is a control parameter. The case $\beta =0$ is
degenerate and requires higher degree nonlinear terms for its description.
We consider only the non-degenerate  case and normalise $\beta$ to 
$\mid\beta\mid =1$ through a suitable scale transformation. The inclusion 
of quadratic terms in the right hand side of eq. (2) does not lead to 
a distinct  bifurcation scenario and so, for the sake of simplicity, 
we omit such terms. 

A typical period-2 orbit of the map (2) is of the form
$$x_n=(-1)^nb_0 \eqno (3a),$$
\noindent with
$$b_0^2 = {\epsilon + q^2 \over \beta}\eqno (3b),$$

\noindent  It follows from eqn.(3b) that a period-2 orbit can  bifurcate
from the fixed point in two ways, and the corresponding bifurcation 
diagrams are presented in 
fig.s 2-3, each type being associated with specific signatures of the 
control parameter $\beta $.  

In each figure the horizontal line depicts the variation of $q$ for fixed 
$\epsilon$, while the distance of each point of the  bifurcating branch from 
the $q$-axis depicts the amplitude of the orbit for the given $q$ and
$\epsilon$. The horizontal q-axis is drawn as a solid (broken) line to 
indicate 
that the symmetric fixed point about which the periodic orbit is 
bifurcating, is
linearly stable (unstable). Similarly, the branch, or portions thereof, 
denoted with
solid lines (broken lines) depict linearly stable  
(unstable) orbits -- anticipating the stability results presented below.

The linear stability analysis of the bifurcating period-2 orbit is 
based on the
variational equation, obtained by linearising eqn.(2), around a
period-2 orbit ${\bar x}_n$~:
$$\xi_{n+2}+\xi_{n-2}-2(q-2)(\xi_{n+1}+\xi_{n-1})+(q^2-4q+6+\epsilon )
\xi_n =3\beta \bar X_n^2 \xi_n ~~~(\xi_n\equiv X_n-\bar X_n)~~~\eqno (4).$$

\noindent In general, for an arbitrary map and for any arbitrary period-two 
orbit, the variational equation would be a linear difference equation with 
period-two coefficients, and its
general solution would be made up of basic solutions of the Floquet form
$$\xi_n=\lambda ^n \zeta_0 \eqno (5),$$
where $\lambda$ is a quasi-multiplier (see, for an explanation,
ref. [17]).  In the present case, eqn.s 3(a,b),(4) imply that the 
period-two coefficients are just constants, but we still assume a solution 
of the form (5) to illustrate the procedure in general. Substituting eqn.(5) 
in (4) and equating the coefficients of $\lambda^n$ on both sides, we get
$$(\lambda+{1\over \lambda }) =-2+q\pm \sqrt{2\epsilon+3q^2} \eqno (6).$$

It is to be noted that eqn.(6) provide us with only one quartet of 
quasi-multipliers, while, in general, there should occur q number of 
quartets for a q-periodic point of a 4D reversible map (see ref. [17]).
This is due to the special form of the variational equation indicated above, 
and the second quartet of quasi-multipliers in the present case can 
be obtained just by multiplying the multipliers belonging to the first 
quartet with $e^{i\pi}$. 

The Floquet multipliers of the period-2 orbits
are $\Lambda_i =\lambda_i^2~~(i=1,..4)$ and can, in principle, correspond 
to  any one 
of the four dispositions in the complex plane shown in fig.4 (these 
dispositions are marked $R1$ through $R4$ for easy reference).  

Based on these results, we find that there arise two distinct 
bifurcation {\it types} ({\it vide} ref. [15]) which we describe below.

\noindent {\it Type I bifurcation~: $\beta >0$.}

A period-2 orbit exists on the subthreshold side, $\epsilon <0$,
for $q>\sqrt{-\epsilon}~~$ (region $P1a$ of parameter space, refer to fig. 1) 
and $q<-\sqrt{-\epsilon}~~$ ($P1b$), and is  unstable (disposition $R3$ 
of fig.4). It merges with the fixed point at
$\epsilon =0, ~q=0$ and, on the superthreshold side $\epsilon >0$ (region $P3$ of fig.1), 
gets detached from the fixed point with the same type of instability ($R3$).

\noindent {\it Type II bifurcation~: $\beta <0$.}

For $\epsilon <0$, a period-2 orbit exists  if 
$-\sqrt{-\epsilon}<q<\sqrt{-\epsilon}$ (region $P2$ of fig.1).
From eqn.(6) it follows that 
the period-2
orbit is unstable(with disposition $R3$ of multipliers) for 
$q<-\sqrt{-{2\epsilon\over 3}}$, unstable (with disposition $R2$) 
for $-\sqrt{-{2\epsilon\over 3}}<q<\sqrt{-{2\epsilon\over 3}}$, and stable 
(disposition $R1$) for $q>\sqrt{-{2\epsilon\over 3}}$.
Further, in this type of bifurcation, there exist no period-2 
orbits for $\epsilon >0$.

\bigskip
\noindent {\it The secondary bifurcation}

We see from above that the period-2 orbit appearing in type II 
bifurcation undergoes a transition from stable $(R1)$ to unstable ($R2$) 
configuration through a {\it secondary bifurcation} at
$$q=q_0\equiv \sqrt{-{2\epsilon\over 3}} \eqno (7).$$

The transition is characterised by the fact that the quasi-multipliers 
of the period-2 orbit collide  
pairwise at $e^{i(\pi\pm \theta_0)}$ 
when $q=q_0$, where
$$4sin^2{\theta_0\over 2}= \sqrt{{2\mid\epsilon\mid \over3}} \eqno (8).$$ 
This `secondary collision' involves the bifurcation of {\it invariant
curves} in the vicinity of the period-2 orbit.

Close to the secondary collision, the nonlinear terms become
relevant, and the deviation $\xi_n=X_n-\bar X_n$ of a typical orbit from
the period-2 orbit under consideration is given by the nonlinear
difference equation
$$\xi_{n+2}+\xi_{n-2}-2(q-2)(\xi_{n+1}+\xi_{n-1})+(q^2-4q+6+\epsilon )\xi_n =
\beta [3\bar X_n^2  \xi_n + 3\bar X_n  \xi_n^2 + \xi_n^3  ] \eqno(9).$$
\noindent and, following [15], we seek
solutions of the form
$$\xi_n =A_n +(B_n e^{in\phi} +B_n^* e^{-in\phi})
+(C_n e^{2in\phi} +C_n^* e^{-2in\phi})+\cdots \eqno(10).$$

\noindent Note that the coefficients in eqn.(10) being period-2 in nature,
the coefficients $A_n$, $B_n$, $C_n,\cdots$ will also be of period 2, say,
$$A_n =a+A(-1)^n \eqno (11a),$$
$$B_n =b+B(-1)^n \eqno (11b),$$
$$C_n =c+C(-1)^n \eqno (11c),$$
$$ .............................$$

\noindent where $a,\cdots ,C,\cdots$ are constants to be determined. In eqn.(10), 
$\phi$ is a rotation angle close to the collision angle $\pi +\theta_0$ 
(this is one of the conjugate angles at which the roots  $\lambda_i $ ($i=1,\cdots,4$) 
collide), say, $$\phi =\pi +\theta_0+\hat\phi \eqno (12),$$
\noindent with $\mid \hat\phi \mid <<\theta_0$. Different values of
$\hat \phi$ would lead to a family of invariant curves organised
in islands around the period-2 orbit.

Substituting eqn.(10) in eqn.(11) and equating the coefficients of 1,
$e^{in\phi}$, $e^{2in\phi}$,.... terms respectively one gets a system of
equations in $A_n$, $B_n$, $C_n, \cdots$ . Substituting eqns.(11 a,b,c) 
in these equations and then using an order-by-order perturbation in 
$\mid q-q_0\mid$ one finds that the
leading term in $B_n$ is the one involving $b$ and, of all the
terms in eqn.(10), this constitutes the dominant contribution. The 
next-to-leading contributions in eqn.(10) are found to arise from terms 
involving $A$ in $A_n$ and $C$ in $C_n$. Truncating the perturbation 
calculation at this order, we get
$$A={6\beta b_0\over (q^2-2\epsilon -3q^2)}b^2 \eqno (13a),$$
$$C={3\beta b_0\over ((q-2+2cos2\phi)^2-2\epsilon -3q^2)}b^2 \eqno (13b),$$
$$b^2={((q-2-2cos\phi)^2-2\epsilon -3q^2) \over \gamma} \eqno (13c),$$
\noindent where
$$\gamma ={36\beta^2 b_0^2\over (q^2-2\epsilon -3q^2)}+
{18\beta^2 b_0^2\over ((q-2+2cos2\phi)^2-2\epsilon -3q^2)}+ 3\beta 
\eqno (14).$$ 

\noindent It is found that $\gamma >0$ and hence there exist two 1-parameter 
families of invariant curves (with associated families of 2-tori) 
around the period-2 points with rotation numbers ${\phi\over 2\pi}$ where
the two families correspond to $\hat \phi>\hat\phi_1$ and 
$\hat \phi<\hat\phi_2$
respectively, with 
$$\hat\phi_{1,2} =  \pm {\sqrt {-2\epsilon /3}\over 2sin\theta_0}
\eqno (15).$$

\noindent The two families merge into a single family as $q$ approaches
$q_0$ from above and then detach from the period-2
orbit as a single one-parameter family of invariant curves (together with
associated 2-tori) for $q<q_0$. 
In accordance with the results of ref.s [9] and [10], the secondary 
bifurcation can be described as a `normal reversible Hopf bifurcation' (an
alternative nomenclature is the Naimark-Sacker bifurcation, see [17]).

Fig.5 shows some of the invariant curves in 2-D projection, obtained on
numerical iterations, on the (a) subthreshold and (b) superthreshold side
(for details, see legend) of the secondary bifurcation. These iterations 
corroborate the results stated above relating to the secondary bifurcation
associated with second order resonant collisions in families of 4D 
reversible maps. While the invariant curves in fig.(5a) belong to 
a so-called `attached' family, those in fig.(5b) are members of what may be 
termed a `detached' family of invariant curves (see sec. 5 below for an 
explanation).

\bigskip
\noindent {\bf IV. THIRD ORDER RESONANT COLLISIONS.}
\bigskip

The 2-parameter family of 4D maps, written in the form of the following 
4th order difference equation, describes a third order resonant collision at 
$q=\epsilon =0 ~$~:
$$x_{n+2} + x_{n-2} -2(q-1)(x_{n+1}+x_{n-1}) + (q^2-2q+3+\epsilon)x_n =
\alpha x_n^2 + \beta x_n^3\eqno(16),$$
\noindent where $\epsilon$ and $q$ are  small bifurcation parameters and, 
once again, the origin is a symmetric fixed point.

It is to be noted that the second degree non-linear term in eqn.(16) 
happens to be 
adequate in describing the bifurcation scenario for the third order resonant 
collision, since third and higher degree terms  are found to lead to no new 
bifurcation types while, for the second order collision, for which the 
linearisation 
of the map at $ q = \epsilon = 0 $ has a distinct Jordan block structure, the 
third degree term is necessary to describe the full scenario. Hence we set 
$\beta = 0$ in the following, though in our numerical iterations a non-zero 
value of $\beta$ has been adopted.

The multipliers of the origin undergo a collision at $e^{\pm ~{2\pi i\over 3}}$ 
on the unit circle when  $q=0$ and $\epsilon =0$ and, for small $ q, 
\epsilon$, bifurcating period-3 orbits exist close to the fixed point. 
A typical bifurcating period-3 orbit is of the form $$\bar x_n = a + 
be^{{2n\pi i\over 3}} +b^*e^{-{2n \pi i \over 3}}\eqno (17),$$ 
\noindent where $a$ and $b$ can be obtained by substituing eqn.(17) in eqn.(16)
and equating the coefficients of the constant term and of 
$e^{{2n\pi i\over 3}}$ respectively.

One can evaluate these constants perturbatively in terms of the 
bifurcation parameters. For the purpose of the present paper where we 
are interested only in  describing the different bifurcation scenarios,
we retain only the terms of the leading order in the bifurcation parameters. 

Such a leading order calculation shows that, for any given $q$ and 
$\epsilon$ there occur, apart from the trivial solution $a=b=0$, two 
distinct non-trivial solutions for b, associated with a 
single solution for a. However, both the solutions are found to correspond 
to the $same$ period-3 orbit when substituted in eqn.(17). A further 
observation is that the bifurcating orbit is independent of the signature 
of $\alpha$, and we can chose $$\alpha > 0$$ with no loss of generality. 
In summary, there occurs only 
one bifurcating period-3 orbit for each point in parameter space, and the 
values of $a,b$, describing this orbit are (with $\alpha$ chosen positive)

$$ b=b^* ={~{\epsilon +q^2} \over \alpha} \eqno(18a),$$
$$a={2\alpha\mid b\mid^2\over (q-3)^2}\eqno (18b).$$

Linear stability analysis of the period-3 orbit can be done in essentially 
the same way as in the period-2 case, the multipliers  
($\Lambda =\lambda^3$, where $\lambda$ is a quasi-multiplier) 
being given by $$(\Lambda +{1\over\Lambda})^2 +A( \Lambda +{1\over
\Lambda}) 
+B=0\eqno (19),$$

\noindent where, up to $O({\epsilon^2})$ terms,
$$A=(-4+6q^2-6\epsilon)+(6\epsilon q-2q^3) +{8\over 3}(\epsilon +q^2)^2
\eqno (20a),$$
$$B=(4-12q^2+12\epsilon)+(4q^3-12\epsilon q)-{291\over 9}(\epsilon +q^2)^2
\eqno (20b).$$

\noindent Thus, the multipliers of the period-3 orbit are obtained from  the
expression $$\Lambda +{1\over\Lambda}=2+3\epsilon-3q^2\pm 6\sqrt{\epsilon^2
+q^4+\epsilon~q^2}\eqno(21).$$

\noindent It is easily seen from eqn.(21) that, for arbitrary choice 
of the parameters $\epsilon$ and $q^2,$ one pair of multipliers of the 
period-3 orbit lies on the unit circle and the other on the real line, 
corresponding to the disposition $R3$ in fig.4. In other words,
the bifurcating orbit is unstable in its domain of existence. 

The bifurcation diagram following from these results is presented in 
fig.6. One finds that, for $\epsilon~<~0$ (the subthreshold side of the 
bifurcation), the period-3 orbit merges with the fixed point 
for $ \mid ~q^2+ \epsilon ~ \mid \rightarrow 0$ while,
for $\epsilon~>~0$ (superthreshold side) the orbit is bounded away from 
the fixed point, the orbit being unstable in both the cases. This diagram 
corresponds to what has been termed a bifurcation of $type~ III$ in the 
context of fourth order resonant collisions in ref.[15] (see next section).

In summary, while the bifurcation scenario for the second order 
resonant collision consists of two types of bifurcation (type I and type II
of sec. III), 
only $one$ type (analogous to type III of ref.[15]) is found to occur in the 
bifurcation scenario associated with the third order collision.
In fig.7 we show the 2D projection of typical trajectories starting close to
the period-3 orbit near the bifurcation, where one finds that the   
trajectories do diverge away from the orbit, in conformity with its 
unstable nature.

\bigskip
\noindent {\bf V. RESONANT COLLISIONS OF ORDER FOUR}
\bigskip

In order to describe the bifurcation of period-4 orbits arising from a 
collision of multipliers at $\pm i~(k=4)$, we consider, as in ref.[15],
the following two-parameter family of maps $A_{q,\epsilon}$, written in the 
form of a 4th order difference equation~:
$$x_{n+2} + x_{n-2} +4q(x_{n+1}+x_{n-1}) + (4q^2+2+4\epsilon)x_n =
\beta (x_n^3+ \gamma x_{n+1}x_{n}x_{n-1})\eqno(22),$$
where $q,~\epsilon$ are small bifurcation parameters and $\beta $
and $\gamma$ are control parameters which will be found in the sequel
to govern the nature or the type of the bifurcation. These parameters will
be assumed to satisfy the nondegeneracy conditions
$$\beta \neq 0 ~~~~~~~and ~~~~~~~~\gamma ^2 \neq 1\eqno(23).$$

\noindent Quadratic terms in eqn.(22) have not been included since they 
are found not to affect the scenario of bifurcations involved.
The origin is once again a symmetric fixed point of 
$A_{q,\epsilon}$, and the multipliers at the origin undergo a collision at 
$\pm i$ at $q=0$ and $\epsilon$=0.

\noindent A typical bifurcating  period-4 orbit of (22) is of the form
$$\bar X_n=(i)^nb_0+(-i)^nb_0^* \eqno (24),$$
\noindent where $b_0$ is obtained by substituting eqn (24) in (22) and 
equating the coefficients of $(i)^n$ and $(-i)^n$.
Two distinct types of bifurcating period-4 orbits
are obtained in this way, which we denote by $O1$ and $O2$ respectively.
$$O1~:~~~~~~~~~~~~~~~~~~~~~~~~~~~~~  b_0=b_0^*=\rho_1 \eqno (25a),$$
$${\it with }~~~~~~~~~~~~~~~~~~~~ \rho_1^2={\epsilon +q^2 \over \beta}
\eqno (25b).$$
$$O2~:~~~~~~~~~~~~~~~~~~~~~~~~~~~~~  b_0=\rho_2 (1+i) \eqno (26a),$$
$$ with ~~~~~~~~~~~~~~~~~~~~ \rho_2^2={\epsilon +
q^2 \over \beta(1-\gamma)} \eqno (26b).$$

\noindent As explained in [15], the two types of orbits have distinct 
symmetry properties under the reversing involutions pertaining to (22).

The linear stability analysis for the bifurcating period-4 orbits 
can be done in the same way as that for period-2 orbit from the
appropriate variational equations. This has been presented in detail in 
ref.[15] and will not be repeated here. When the stability results of [15] 
are combined with eqn.s (25a) through (26b), one finds that the bifurcation 
scenario associated with the fourth order resonance involves $three$ bifurcation 
types. We designate these as bifurcations of type I, II, and III,
corresponding respectively to bifurcation diagrams of fig.s 8, 9, and 10. 
A brief description of these types, corresponding to different sets of values 
of the control parameters $\beta$ and $\gamma$, is as follows.

\noindent Type I~: $\beta >0$, $\gamma <1$

For $\gamma <-1$, on the subthreshold side ($\epsilon <0$), both 
the bifurcating branches of period-4 orbits ($O1$ and $O2$) exist for 
$q>\sqrt{-\epsilon}$ and $q<-\sqrt{-\epsilon}$ (i.e., in regions $P1a$ 
and $P1b$ 
of the parameter space, refer to fig.1). In this situation $O1$ is linearly 
stable (disposition $R1$ of fig. 4) and $O2$ is unstable (disposition $R3$) 
(fig.8a). For $\epsilon =0$, they merge together at the origin
at $q=0$(fig.8b) and for $\epsilon >0$, they get detached from the origin 
with their respective stability types remaining unchanged (fig.8c). The 
situation  remains essentially the same for $-1<\gamma<1$, with 
only the stability types getting interchanged (disposition $R3$ for 
$O1$ and $R1$ for $O2$).

\noindent Type II~: $\beta <0$, $\gamma <1$

For $\epsilon <0$,  two branches exist simultaneously for
$-\sqrt{-\epsilon}<q<\sqrt{-\epsilon}$, and are globally connected (as the 
second bifurcation parameter $q$ is varied), getting annihilated 
simultaneously as $\epsilon$ becomes positive.

\noindent Here we encounter once again the interesting situation that 
one of the two branches undergoes a change of stability at some critical 
value of $q$ (for fixed $\epsilon <0$) through a collision of Floquet 
multipliers on the unit circle near +1 (Fig.9). Thus, for $\gamma <0$, 
the branch $O2$ is unstable (disposition $R3$) 
throughout the range $-\sqrt{\mid\epsilon \mid}<q<\sqrt{\mid\epsilon \mid}$
while, for the $O1$ solution, there occurs a transition from linearly stable
(disposition $R1$) to unstable(disposition $R2$) configuration as $q^2$ 
crosses the threshold

$$q_1^2=-\epsilon {1-\gamma+2\sqrt{2}\sqrt{-1-\gamma}\over
5-\gamma -2\sqrt{2}\sqrt{-1-\gamma}}\eqno (27)$$
from above, due to a `small angle' collision (see [19] for an explanation) 
of Floquet multipliers. As $q^2$
is made to decrease further, a second transition occurs at
$$q^2=q_2^2=-\epsilon {1-\gamma-2\sqrt{2}\sqrt{-1-\gamma}\over
5-\gamma -2\sqrt{2}\sqrt{-1-\gamma}}\eqno(28),$$
\noindent the disposition of the mutipliers changing over to $R4$ which, 
however, does not change the unstable character of the branch (this 
transition has not been shown in fig. 9).
For $0<\gamma <1$, on the other hand, the $O1$ orbit is 
unstable (disposition $R3$) throughout
the range $-\sqrt{\mid\epsilon \mid}<q<\sqrt{\mid\epsilon \mid}$, while for
the $O2$ branch there occurs a transition from linearly stable($R1$) to 
unstable ($R2$) configuration as $q^2$ crosses the value
$$q_3^2=-\epsilon {1+\sqrt{1-\gamma ^2}\over 2-\gamma +\sqrt{1-\gamma^2}}
\eqno (29),$$ from above, due to a small angle collision of the multipliers.
As $q^2$ is made to decrease further, a second transition occurs as in 
the case of the $O1$ branch above.

\noindent Type III~: $\gamma >1.$

In this case one branch ($O1$ for $\beta >0$) exists for 
$q>\sqrt{-\epsilon}$ and $q<~-\sqrt{-\epsilon}$ (regions $P1a$ and $P1b$ of 
fig. 1) and the other branch ($O2$) exists for $-\sqrt{-\epsilon}<q
<\sqrt{-\epsilon}$ (Fig.10a). Each of the branches is unstable 
(disposition $R3$) in its domain of existence. For $\epsilon >0$ the branch 
$O1$ gets detached from the origin and $O2$ is annihilated (Fig.10c). 
The branches get interchanged if $\beta < 0$.

Thus, only for the type II collision near $\pm i$ (i.e., for $\beta <0$ 
and $\gamma <1$ ), there occurs a transition of a period-4 branch from 
linearly stable (disposition $R1$) to an unstable ($R2$) configuration 
through a small angle secondary collision. This is the co-dimension 1 
collision discussed in ref.s [9,10]) and so the full co-dimension 2 
collision studied in [19] is not relevant here.  The transition in the 
disposition of multipliers from $R2$ 
to $R4$ referred to above again involves a co-dimension 1 collision, and
will not be discussed here (see, e.g., [18], section 36).

The phase space dynamics in the vicinity of the secondary 
collision can once again be investigated by going beyond the linear terms. 
The analysis is essentially similar to the one outlined in case of the
second order resonant collision and has been presented in detail in
ref. [15]. The principal result in this respect is that {\it the secondary 
bifurcation is superthreshold or `normal' in character} (see [9,10] for an 
explanation). 

One can proceed further from this result and construct 
perturbatively the `islands' of invariant curves surrounding the period-4 
points in the vicinity of the secondary collision.
Thus, on the subthreshold side of the bifurcation, each linearly stable 
period-4 point is surrounded by a family of invariant curves (forming an 
island, there being four such islands for the period-4 orbit under 
consideration), members of 
the family passing arbitrarily close to the period-4 point for rotation 
angles chosen sufficiently close to $2\pi \over 4$. This family persists on 
the superthreshold side of the bifurcation, but now detached from the 
period-4 point under consideration, i.e., there is now a lower bound to the 
distance of orbits belonging to this family from the period-4 point. These 
two types of families, with zero and non-zero lower bounds to distances from  
the fixed point or periodic point under consideration will be termed 
`attached' and `detached' families respectively.

{\it Attached} families of invariant curves are commonly found to occur in 
reversible mappings around linearly stable fixed or periodic points (see ref.  
[15]), while there exists no standard result in the literature 
concerning {\it detached} families. 
As described in ref.s [10,11,15], such detached families are to be observed 
in the vicinities of `normal' reversible Hopf bifurcations. We have already 
come across attached and detached families of invariant curves around 
period-2 points in sec. III.

Figures 11 and 12 represent results of numerical iterations with the 
family of
maps $A_{q,\epsilon}$ (eqn. (22)) close to the secondary bifurcation.
Figure 11a shows three invariant curves belonging to an attached  
family forming an
island around one of the period-4 points of $O1$ on the subthreshold
side of the bifurcation (for details, see legend) in a two-dimensional
projection. Figure 11(b) on the other hand shows three invariant curves
belonging to a detached family on the superthreshold side of the secondary 
bifurcation around the $O1$
orbit. Figures 12(a,b) show members belonging to corresponding families,
on the subthreshold and superthreshold sides respectively, for the $O2$ 
branch (again,see legend, for details of each figure).

Two-dimensional projections of trajectories initiated very close to the 
two types of period-4 orbits ($O1$ and $O2$), on subthreshold and 
superthreshold sides of each of the three types of bifurcation 
have been presentend in fig.s 
13, 14, and 15 respectively. In each figure the persistence of the 
trajectory signifies stability of that orbit, whereas 
a precipitate departure away from the orbit signifies its instability. 
Thus, fig. 13(a,b) represent 2D projections of trajectories initiated
very close to the two types of period-4 orbits in type I bifurcation, where
one finds that the trajectory around O1 persists near it forming an island, 
while one initiated near O2 departs away from it, following a path close to 
the separatrix. 

Similarly, fig. 14(a-b) depict trajectories initiated near the O1 
and O2 orbits in type II bifurcation, now close to the {\it secondary} 
bifurcation. While fig. 14(a) corresponds to the subthreshold side of the 
secondary bifurcation, in which the trajectory initiated close to O1 forms 
an island and the one initiated close to O2 makes a circuit in the 
vicinity of the separatrix, fig. 14(b) corresponds to the {\it 
superthreshold} side, where trajectories are seen to depart away from {\it 
both} O1 and O2 orbits. 

Fig. 15 presents similar trajectories for type III bifurcation in the 
fourth order resonant collision. As seen from fig. 
10, there exists only one type of orbit ($O1$ {\it or} $O2$) for each 
specification of parameter values ($q,~\epsilon$) on either side 
of the bifurcation ($\epsilon < 0, \epsilon > 0$), and that orbit is 
unstable. Fig. 15(a, b) show trajectories initiated close to the $O2$ and 
$O1$ orbits respectively for $\epsilon < 0$, and for two distinct values of 
$q$, and in each case the trajectory is seen to diverge away from the 
period-4 orbit concerned. Fig. 15(c), on the other hand depicts a similar 
trajectory for $\epsilon > 0$, where only the $O1$ orbit, now 
{\it detached} from the fixed point, survives.

\bigskip
\noindent {\bf VI. HIGHER ORDER RESONANT COLLISIONS.}
\bigskip

\leftline {~}
\noindent {\it Collisions of order five.}

We consider fifth order resonant collisions with reference to a family of 
maps $A_{q,\epsilon}$ written in the form of a fourth order difference 
equation,
$$X_{n+2} + X_{n-2} -4cos\phi_0(X_{n+1}+X_{n-1}) + (4cos^2\phi_0+2+\epsilon)X_n =
\alpha X_n^2 + \beta X_n^3\eqno (30),$$
\noindent where now $cos\phi_0=cos{2\pi\over 5}+q/2 $, $\epsilon$  
and $q$ being small bifurcation parameters as in the previous sections.

The multipliers of the origin undergo a
collision at $e^{{2\pi i\over 5}}$ at $q=0$ and $\epsilon =0$. A typical
bifurcating period-5 orbit is of the form
$$\bar x_n = a + be^{{2n\pi i\over 5}} +b^*e^{{-2n\pi i\over 5}}
+ ce^{{4n\pi i\over 5}} +c^*e^{{-4n\pi i\over 5}}\eqno (31),$$
\noindent where $a,~b,~b^*,~c,~c^*$ can be obtained  perturbatively by 
substitution in eq. (30).

In the leading order in $\epsilon$ and $q^2$, one finds
$$a_0 \approx {2\alpha{\mid b\mid}^2 \over \epsilon_0}\eqno (32a),$$
$$c \approx {\alpha b^2 \over \epsilon_2}\eqno (32b),$$
$$c^* \approx {\alpha {b^*}^2 \over \epsilon_2}\eqno (32c),$$
\noindent and
$$\mid b \mid~=~\rho_0 \eqno (32d),$$
\noindent with
$$\rho_0^2={\epsilon_1 \over \gamma} \eqno(33a),$$
\noindent where 
$$\gamma =2\alpha^2({2\over\epsilon_0} +{1\over\epsilon_2}) +3\beta \eqno
(33b), $$
the parameters $\epsilon_0,~\epsilon_1$, and $\epsilon_2$ being given by
$$ \epsilon_0= \epsilon +4(cos\phi_0 -1)^2 \eqno (34a),$$
$$\epsilon_1= \epsilon +q^2 \eqno (34b),$$
$$\epsilon_2= \epsilon +4(cos\phi_0 -cos{4\pi\over 5})^2 \eqno (34c),$$
respectively.

These equations tell us that, in the leading order of perturbation, there 
is only one period five orbit for any given value of the parameters 
$\epsilon$ and $q$. 
However, on going over to the next order of perturbation (i.e., 
to terms of the second degree in $\epsilon$ and $q$)
one finds that there exist, in fact, {\it two} distinct types of period-5 
orbits, which we designate as $O1$ and $O2$ respectively, distinguished by 
the value of the coefficient $b$ in eq. (31)~:

$$O1~:~~~~~~~~~~~~~~~~~~~~~~~~~~~~~  b=b^*=\rho_a \eqno (35a),$$
$${\it with }~~~~~~~~~~~~~~~~~~~~ \rho_a=\rho_0+\rho_{1a}\eqno (35b).$$
$$O2~:~~~~~~~~~~~~~~~~~~~~~~~~~~~~~  b=\rho_b e^{{\pi i\over 5}}
\eqno (36a),$$
$${\it with }~~~~~~~~~~~~~~~~~~~~ \rho_b=\rho_0+\rho_{1b}\eqno (36b).$$

\noindent Where $\rho_{1i}~ (i~=~a, b)$ is given by

$$\rho_{1i}=-{\rho_0^4 \chi cos4\phi_i \over 2\epsilon cos\phi_i +
4\chi \rho_0^3cos4\phi_i}~~~~(i=a,b)\eqno (37), $$

\noindent with 

$$ \phi_a~=~0,~~\phi_b~= ~ {\pi \over 5} \eqno 38(a,b),$$
and $$\chi = {\alpha^3\over \epsilon_2^2}+{3\alpha\beta\over\epsilon_2} 
\eqno (39).$$

\noindent In other words, there is a degeneracy between the two
types of period-5 orbits in the leading order of perturbation, which is   
removed in the next higher order. It is also apparent from eqn. (39) that the 
quadratic term in eqn. (30) is important in this removal of degeneracy. 

\noindent Having obtained the perturbative expressions for the period-5 
orbits it is a routine, if tedious, affair to follow trajectories initiated 
close to the orbits for various sets of parameter values, scanning across the 
parameter space. Perturbation calculations of the type resorted to in the 
case of lower order resonant collisions present increasing magnitudes of 
difficulty as one comes across resonant collisions of order $5$ and above.
Hence, in absence of a 
perturbative stability analysis of the orbits and a perturbative 
construction of the families of invariant curves close to these orbits,
we invoke the approach of numerical construction of trajectories close to 
the periodic orbits so as to have an idea of the bifurcation {\it scenarios} 
involved.

In the present instance, such an approach does seem to lead to a reliable 
picture, which we summarise as follows~:

\noindent Period-5 orbits can bifurcate from a fixed point in a fifth 
order resonant collision in either of two ways that resemble the type I and 
type II bifurcations in a fourth order resonant collision. Thus, in 
type I bifurcation there exists a stable and an unstable orbit on either side 
of the bifurcation ($\epsilon < 0$, and $\epsilon > 0$), while in type II 
bifurcation, a pair of orbits exists only on the subthreshold side 
$\epsilon < 0$, there being no period-5 orbit on the superthreshold side. 
Of the members of the pair bifurcating from the fixed point for $\epsilon
 < 0$, one is always unstable, while the other undergoes a secondary 
bifurcation which is `normal' in nature (see sec. V).

Fig. 16 presents trajectories initiated close to the period-5 
orbits in a type I bifurcation for $\epsilon < 0$ (fig. 16a) and $\epsilon 
> 0$ (fig. 16b). One does find from the figure that the trajectory near one 
of the orbits belongs to an island of invariant curves, signifying that the 
orbit concerned is stable, while the trajectory near the other orbit moves 
away following a separatrix loop, and that essentially the same picture 
persists on the two sides of the bifurcation.

Fig. 17, on the other hand, depicts a {secondary bifurcation } 
associated with a type II bifurcation where one finds that trajectories 
on the subthreshold side of the secondary bifurcation (fig. 17a) are similar 
to those in the type I bifurcation while, on the superthreshold side 
(fig. 17b), trajectories move away from {\it both} the orbits, signifying 
that they are now unstable.

\leftline {~}
\noindent {\it Collisions of order six.}
\bigskip
Bifurcations of period-6 orbits can be studied with the family 
of maps ($A_{q, \epsilon }$) written in the form
$$X_{n+2} + X_{n-2} -2(1+q)(X_{n+1}+X_{n-1})+ (q^2+2q+3+\epsilon )X_n =
\alpha X_n^2 + \beta X_n^3 \eqno (40),$$

\noindent where, as before, $\epsilon$ and $q$ are  small bifurcation 
parameters, and $\alpha, \beta$ are control parameters.

As seen from eq. (40) the multipliers of the origin undergo a
collision at $e^{{2\pi i\over 6}}$ at $q=0$ and $\epsilon =0$. A typical
bifurcating period-6 orbit is of the form
$$\bar x_n = a + d(-1)^n + be^{{2n\pi i\over 6}} +b^*e^{{-2n\pi i\over 6}}
+ ce^{{4n\pi i\over 6}} +c^*e^{{-4n\pi i\over 6}} \eqno (41),$$
\noindent where $a,~d,~b,~b^*,~c,~c^*$ can be obtained  perturbatively.
In the leading order of perturbation in $\epsilon$ and $q^2$ one finds
$$a_0 \approx {2\alpha{\mid b\mid}^2 \over \epsilon_0} \eqno (42a),$$
$$c_0 \approx {\alpha b^2 \over \epsilon_2} \eqno (42b),$$
$$c_0^* \approx {\alpha {b^*}^2 \over \epsilon_2} \eqno (42c),$$
$$d_0 \approx {\delta(b^3+{b^*}^3 )\over \epsilon_3} \eqno (42d),$$
\noindent and
$$\mid b \mid~=~\rho_0 \eqno (42e),$$
\noindent with
$$\rho_0^2={\epsilon_1 \over \gamma} \eqno (43a),$$
\noindent where 
$$\gamma =2\alpha^2({2\over\epsilon_0} +{1\over\epsilon_2}) +3\beta 
\eqno (43b), $$
$$ \epsilon_0= \epsilon +(q -1)^2 \eqno (43c),$$
$$\epsilon_1= \epsilon +q^2 \eqno (43d),$$
$$ \epsilon_2= \epsilon +(q +2)^2 \eqno (43e),$$
$$ \epsilon_3= \epsilon +(q +3)^2 \eqno (43f),$$
$$\delta ={2\alpha^2\over\epsilon_2 }+\beta \eqno (43g).$$

Thus, there exists only one bifurcating orbit in the leading order of 
perturbation which, however is to be interpreted as just a degeneracy between 
distinct orbits that is lifted in the higher orders. Indeed, in the next 
order of perturbation, one finds that as in the case of a resonant collision
of order five there exist two distinct types of period-6 orbits, which we 
describe as follows~:

$$O1~:~~~~~~~~~~~~~~~~~~~~~~~~~~~~~  b=b^*=\rho_a \eqno (44a),$$
$${\it with }~~~~~~~~~~~~~~~~~~~~ \rho_a=\rho_0+\rho_{1a} \eqno (44b),$$
$$O2~:~~~~~~~~~~~~~~~~~~~~~~~~~~~~~  b=\rho_b e^{{\pi i\over 6}} \eqno (45a),$$
$${\it with }~~~~~~~~~~~~~~~~~~~~ \rho_b=\rho_0+\rho_{1b} \eqno (45b),$$
\noindent $\rho_{1i}~ (i~=~a,b)$ being given by
$$\rho_{1i}=-{\rho_0^5(\chi_1 cos\phi_i +\chi_2 cos5\phi_i )\over 
2\epsilon cos\phi_i+5\rho_0^4(\chi_1 cos\phi_i +\chi_2 cos5\phi_i )} 
\eqno (46), $$
\noindent where
$$\phi_a~=~0,~~\phi_b~={\pi} \over ~6 ~\eqno (47 a,b),$$
$$\chi_1 = {2\alpha^2\delta\over\epsilon_2\epsilon_3}+{3\beta\delta\over\epsilon_3}
+{6\alpha^2\beta\over\epsilon_2^2}+{12\alpha^2\beta\over\epsilon_0^2}
+{12\alpha^2\beta\over\epsilon_0\epsilon_2} \eqno (47c),$$
$$\chi_2 = {2\alpha^2\delta\over\epsilon_2\epsilon_3}+{3\beta\delta\over\epsilon_3}
+{3\alpha^2\beta\over\epsilon_2^2} \eqno (47d).$$

\noindent Once again, having obtained the perturbative expressions for the 
period-6 orbits, the dynamics in phase space close to these orbits can be 
studied through numerical iterations in order to construct the bifurcation 
scenario. The result emerging from such an exercise is~:

\noindent {\it The scenario in the sixth order resonant collision is 
analogous to that in the fifth order collision, consisting of two types 
of bifurcation resembling in turn the type I and type II bifurcations 
associated with the fourth order collision. In particular, the type II
bifurcation involves a secondary collision whereby one of the two period-6 
orbits undergoes a transition from the linear stablility to  
instability.}

Evidence for the two types of bifurcations in the family of maps 
given by eqn.(40) is presented in the form of fig.s 18 and 19, corresponding 
in significance to fig.s 16 and 17 respectively for the two types of 
bifurcation in the fifth order resonant collision (see legends for details).

These observations based on numerical iterations, together with the 
results of the previous sections lead to a conjecture concerning resonant 
collisions in families of reversible mappings presented in the next section.

\bigskip
\noindent {\bf VII. SUMMARY~: A CONJECTURE.}
\bigskip

In this paper we have considered resonant collisions of the 
multipliers at a symmetric fixed point of families of 4D reversible maps 
specified as in sec. II. A resonant collision has 
been defined in sec. I as a collision of multipliers on the unit circle at
angles $\pm {2\pi\over k}$ ($k$ positive integer). The case $k~=~1$ involves 
the bifurcation of {\it fixed points}, as distinct from the cases $k~>~1$ 
where periodic orbits are found to bifurcate from the fixed point under 
consideration, and has been investigated from a numerical point of view in 
ref.[19]. The present paper addresses resonant collisions with $k~>~1$.
The families of maps we have considered for different specified 
values of k, while conforming to the criteria mentioned in sec. II, have 
all been chosen to be of the so-called de Vogelaere type (see, e.g., ref.
[9]), these being relatively easy to handle analytically and numerically 
(eqn.22, corresponding to $k~=~4$, is an exception since the de Vogelaere form
does not lead to the complete scenario in this case). 
                                                                        
\noindent Collecting the results obtained from perturbative calculations and 
numerical iterations we formulate the following conjecture~:

\noindent {\it There exists a characteristic bifurcation scenario for each 
of the cases $k = 2, 3, 4$, and a fourth and distinct scenario for $k~>~4$.

The scenario for k = 2 involves two bifurcation types as depicted in fig.s 
2 and 3, and in each of these there exists only one bifurcating period-2 
orbit. This orbit undergoes a secondary bifurcation at a particular value of 
q ($0 < q < \sqrt (-\epsilon)$), with attached and detached families of 
invariant curves on the subthreshold and superthreshold sides of the 
bifurcation respectively.

There is only one bifurcation type involved in the scenario for k = 3 
(fig. 6), the associated period-3 orbit being unstable on either side of the 
bifurcation. This bifurcation is analogous to one of the three types 
involved in the scenario for k~=~4.

Three bifurcation types are involved in the scenario for k = 4 (fig.s 8, 9, 
10). We have termed these type I, type II, and type III bifurcations 
respectively. In one of these (type II) there occurs a secondary bifurcation 
of one of the two associated period-4 orbits at {\it two} symmetrically located 
values of $q$ for any given $\epsilon < 0$.

The scenario for any $k > 4$ involves two bifurcation types analogous 
respectively to types I and II for $k = 4$ (fig.s 8, 9).}

The bifurcating periodic orbits are associated with families of invariant 
curves which may be of either `attached' or `detached' type (see sec. V for 
explanation) corresponding to whether the orbit under consideration is 
linearly stable or is an unstable orbit on the superthreshold side of a 
secondary bifurcation.  Our perturbative and numerical approach
gives us a good idea of these families sufficiently close to the resonant 
collisions. 

Additionally, one can talk of 2-dim {\it tori} in the immediate 
neighbourhoods of the bifurcating orbits. Analogous to the families of 
invariant curves, these tori can also be of the `attached' and `detached' 
types. Thus, each of fig.s 20(a,b) presents a two-dimensional projection
of a torus around a period-4 orbit, respectively on the subthreshold and 
superthreshold sides of the secondary bifurcation in a resonant collision 
leading to a type II bifurcation.

In other words, it is possible to have considerable information of the 
dynamics in phase space close to a resonant collision of multipliers of a 
symmetric fixed point of a family of 
4D reversible mappings from the perturbative approach coupled with 
numerical iterations of the type presented in this paper. One hopes that this 
will help in the formulation of rigorous results in this largely uncharted 
area [20].
\bigskip
\noindent {\bf Acknowledgement.} We thank M. B. Sevryuk for suggesting the 
present area of investigation to one of us (A.L).

Several of the figures presented in this
paper have been taken from a previous paper of ours (ref.[15]) and a few 
others from the Ph. D. thesis of one of us (A.B).

\vfil\eject
\leftline {\bf References.}

\item{[1]} M.G. Krein, Amer. Math. Soc. Trans. Ser. 2, {\bf 120} (1955), 1.
\item{[2]} A.Deprit and J.Henrard, Adv. Astron. Astrophys {\bf 6} (1968), 1.
\item{[3]} J. van der Meer, The Hamiltonian Hopf Bifurcation, Lecture 
Notes in Math, {\bf 1160}, (ed.) A. Dold and B. Eckmann (Springer-Verlag, 
Berlin, 1985)
\item{[4]} K.R. Meyer and G.R. Hall, Introduction to Hamiltonian Dynamical
 Systems and the N-body problem; Springer-Verlag, Berlin, 1992.
\item{[5]} T.J.Bridges, Math. Proc. Camb. Phil. Soc. {\bf 108} (1990), 575.
\item{[6]} A.Lahiri, to be communicated.
\item{[7]} V.I.Arnold and M.B.Sevryuk, Oscillations and Bifurcations in 
Reversible Systems, in {\it Non-linear Phenomena in Plasma Physics and 
Hydrodynamics}, ed. R.Z.Sagdeev (English translation by Valerii
Ilyushchenko), Mir Publishers, Moscow, 1986);
\item{[8]} M.B.Sevryuk, {\it Reversible Systems}, Lecture Notes in Mathematics, 
{\bf 1211} ed. A.Dold and B. Eckmann, Springer Verlag, Berlin, 1986.
\item{[9]}  M.B. Sevryuk and A. Lahiri, Phys. Letts. A {\bf 154} (1991), 104.
\item{[10]}  T.K. Roy and A. Lahiri, Phys. Rev. A {\bf 44} (1991), 4937.
\item{[11]} A. Bhowal, Reversible Hopf Bifurcation in a Family of Four 
Dimensional Maps  (Ph.D Thesis, Calcutta Univ, 1996).
\item{[12]} T. J. Bridges, R.S.MacKay, and R.H.Cushman, Dynamics near an Irrational 
Collision of Eigenvalues for Symplectic Maps(1993), Warwick Reprints 10/1993.
\item{[13]} T. J. Bridges and J. E. Furter, {\it Singularity Theory and 
Equivariant Symplectic Maps}, Springer-Verlag, Berlin, 1993.
\item{[14]} T. J. Bridges, J. E. Furter and A. Lahiri, Collision of Multipliers at
 $\pm i$ in Reversible-Symplectic Maps, Preprint, University of Warwick (1991).
\item{[15]} A.Lahiri, A.Bhowal and T.K.Roy, Physica D, {\bf 85} (1995), 10.
\item{[16]} D. Pfenninger, 
Astron. and Astrophys. {\bf 150} (1985),  97~;
Astron. and Astrophys. {\bf 180} (1987), 79.
\item{[17]} A.Lahiri, A.Bhowal, T.K.Roy and M.B.Sevryuk, Physica-D {\bf 63}
(1993), 99.
\item{[18]} V. I. Arnold~, {\it Geometrical Methods in the Theory of Ordinary 
Differential Equations}, Second Edition, Springer-Verlag, New York (1988).
\item{[19]} A.Bhowal, T.K.Roy and A.Lahiri, Phys Rev E {\bf 47} (1993),3932.
\item{[20]} It appears (T. J. Bridges, private communications) that a set of 
rigorous results have been arrived at by J. Furter and by I. Hoveijn in the 
context of reversible-symplectic maps.
\vfil\eject
\leftline{ FIGURE CAPTIONS.}

\noindent {\bf Fig. 1} ~~Regions in the ($\epsilon -q$) parameter sapce
divided by the curves $(i) ~\epsilon=0$ and (ii)~$q^2 +\epsilon =0$.

\bigskip

\noindent {\bf Fig. 2} ~~Schematic bifurcation diagram for Type I bifurcation 
($\beta>0$) of a period-2 orbit around a symmetric fixed point for 
(a) $\epsilon <0$, (b) $\epsilon =0$, (c) $\epsilon >0$. For explanation, 
see text, Sec. III.

\bigskip

\noindent {\bf Fig. 3} ~~Schematic bifurcation diagram for Type II bifurcation 
($\beta<0$) of a period-2 orbit around a symmetric fixed point for 
$\epsilon <0$. For explanation, see text. Note the occurence of the secondary 
bifurcation. The superthreshold side ($\epsilon >0$) is not shown in the 
figure since there exists no period-2 orbit on this side.

\bigskip

\noindent {\bf Fig. 4} ~~Different possible disposition of the Floquet
multipliers at a symmetric fixed point or periodic orbit for 4D reversible maps.
For explanation, see text, Sec III.

\bigskip

\noindent {\bf Fig. 5} ~~(a)~2-dim projection of a set of invariant curves around one of
the period-2 points in second order resonant collision (type II bifurcation)
on the subthreshold
side of the secondary bifurcation, with
$\epsilon =-10^{-5}$, $\beta=-1$ and
$q=2.585\times 10^{-3}$, corresponding to  $\hat\phi=2.17\times 10^{-3}$,
$2.37\times 10^{-3}$ 
and $2.57\times 10^{-3}$ respectively.
(b)Projection of a set of invariant curves around the period-2 point 
on the superthreshold side of the secondary bifurcation, with
$q=-2.58\times 10^{-3}$ for $\hat\phi= 1\times 10^{-4}$,
and $8\times 10^{-5}$ respectively, other parameters
being the same as in (a).

\bigskip

\noindent {\bf Fig. 6} ~~Schematic bifurcation diagram of period-3 bifurcation~;
(a) $\epsilon <0 $, (b) $\epsilon =0$, (c) $\epsilon >0$.

\bigskip

\noindent {\bf Fig. 7} ~~2-dim projection of a trajectory  initiated 
close to a period-3 point near the resonant collision with 
$\beta=1$, $\alpha =1$  corresponding to
(a) $\epsilon =-1\times 10^{-4}$, $q =8\times 10^{-3}$ and
$\delta=1\times 10^{-9}$,
and (b) $\epsilon =1\times 10^{-5}$,  $q =8\times 10^{-3}$ and
$\delta=1\times 10^{-9}$; here and in the subsequent figures
the parameter $\delta$  specifies the initial condition for the
trajectory;
we first calculate  period-3 orbit ($\bar X_n$) from eqn.(17) and 
then take the initial condition of the trajectory as 
$X_n=\bar X_n +n(-1)^n \delta ~ (n=1,2,3,4)$.

\bigskip

\noindent {\bf Fig. 8} ~~Schematic bifurcation diagram for Type I bifurcation
($\beta>0,~\gamma<1$) of period-4 orbits around a symmetric fixed point for 
(a) $\epsilon <0$, (b) $\epsilon =0$, (c) $\epsilon >0$. The labels
$O1$ and $O2$ are shown for $\gamma <0$ while for $0<\gamma <1$ they should
be interchanged.  For explanation, see text, sec.VI.

\bigskip

\noindent {\bf Fig. 9} ~~Schematic bifurcation diagram for Type II bifurcation
($\beta<0,~\gamma<1$). Only the subthreshold situation  ($\epsilon <0$) is
shown. The orbits shrink to the fixed point at $\epsilon =0$, and cease to
exist for $\epsilon >0$. The labels
$O1$ and $O2$ are shown for $\gamma <0$ while for $0<\gamma <1$ they should
be interchanged.

\bigskip

\noindent {\bf Fig. 10} ~~Schematic bifurcation diagram of Type-III bifurcation
($\gamma>1$); (a) $\epsilon <0 $,
(b) $\epsilon =0$, (c) $\epsilon >0$. The labels
$O1$ and $O2$ are shown for $\beta >0$ while for $\beta <0$ they should
be interchanged; see text, sec.VI.

\bigskip

\noindent {\bf Fig. 11} ~~(a)~2-dim projection of a set of invariant curves around one of
the period-4 points of the $O1$ orbit on the subthreshold side of the secondary
 bifurcation, with $\epsilon =-10^{-5}$, $\beta=-1$, $\gamma=-3.02$ and
$q=-2.585\times 10^{-3}$ corresponding to $\hat\phi =1.173\times 10^{-4}$,
$1.178\times 10^{-4}$ and $1.183\times 10^{-4}$ respectively.
(b)~Projection of a set of invariant curves around one of the
period-4 points of the $01$ orbit
on the superthreshold side of the secondary bifurcation, with
$q=-2.5835\times 10^{-3}$ for $\hat\phi= 1\times 10^{-5}$,
$3\times 10^{-5}$ and $5\times 10^{-5}$ respectively, other parameters
being the same as in (a).

\bigskip

\noindent {\bf Fig. 12} ~~(a)~2-dim projection of a set of invariant curves around one of
the period-4 points of the $O2$ orbit on the subthreshold side of the secondary
bifurcation, with $\epsilon =-10^{-5}$, $\beta=-1$, $\gamma=-10^{-2}$ and
$q=-2.578\times 10^{-3}$ corresponding to  $\hat\phi=6.99\times 10^{-5}$,
$7.02\times 10^{-5}$ and $7.05\times 10^{-5}$ respectively.
(b)~Projection of a set of invariant curves around one of
the period-4 points of the $O2$ orbit
on the superthreshold side of the secondary bifurcation, with
$q=-2.577\times 10^{-3}$ for  $\hat\phi =1\times 10^{-5}$,
$2.2\times 10^{-5}$ and $3.4\times 10^{-5}$ respectively, other parameters
being the same as in (a).

\bigskip

\noindent {\bf Fig. 13} ~~2-dim projection of two trajectories  starting from
close to the period-4 orbits ($O1$ and $O2$) in type I bifurcation near the
resonant collision with $\beta=1$,
$\gamma =0$ and $q=2\times 10^{-2}$, corresponding to (a) $\epsilon =-1\times 10^{-4}$,
$\delta=1\times 10^{-4}$ for $O2$ and $\delta=0$ for $O1$ orbit and
(b) $\epsilon =1\times 10^{-4}$,
$\delta=1\times 10^{-4}$ for $O2$ and $\delta=0$ for $O1$ orbit;
we first calculate  period-4 orbit ($\bar X_n$) from eqn.(24) and 
then take the initial condition of the trajectory as $X_n=\bar X_n +\delta$ 
($\delta$ being taken  nonzero only for $n=4$).

\bigskip

\noindent {\bf Fig. 14} ~~2-dim projection of two trajectories  starting from
close to the period-4 orbits ($O1$ and $O2$) in type II bifurcation near the
$secondary$ bifurcation  with $\beta=-1$, $\gamma =0$ and 
$\epsilon =-1\times 10^{-4}$, corresponding to (a) $q =9.5\times 10^{-3}$,
$\delta=5\times 10^{-6}$ for $O2$ and $\delta=1\times 10^{-6}$ for $O1$ orbit 
and (b) $q =8.1\times 10^{-3}$,
$\delta=1\times 10^{-6}$ for $O2$ and $\delta=1\times 10^{-6}$ for $O1$ orbit~;
we first calculate  period-4 orbit ($\bar X_n$) from eqn.(24) and 
then take the initial condition of the trajectory as $X_n=\bar X_n +\delta$ 
($\delta$ being taken nonzero only for $n=4$).

\bigskip

\noindent {\bf Fig. 15} ~~2-dim projection of a trajectory  starting from
close to the period-4 orbit ($O1$ or $O2$) in type III bifurcation near the
bifurcation  point with $\beta=1$, $\gamma =2$  corresponding to
(a) $\epsilon =-1\times 10^{-4}$,  $q =-9\times 10^{-3}$ and
$\delta=2.5\times 10^{-6}$ for $O2$,
(b) $\epsilon =-1\times 10^{-4}$,  $q =-1.2\times 10^{-2}$ and
$\delta=-1\times 10^{-6}$ for $O1$  
and (c) $\epsilon =1\times 10^{-5}$,  $q =-1.2\times 10^{-2}$ and
$\delta=-1\times 10^{-8}$ for $O1$ orbit; 
we first calculate  period-4 orbit ($\bar X_n$) from eqn.(24) and 
then take the initial condition of the trajectory as 
$X_n=\bar X_n +n(-1)^n\delta cos(n{\pi\over 4})$.

\bigskip

\noindent {\bf Fig. 16} ~~2-dim projection of two trajectories  starting from
close to the period-5 orbits ($O1$ and $O2$) in type I bifurcation near the
bifurcation point with $\beta=-.5$,
$\alpha =1$ and $q=6\times 10^{-2}$ corresponding to (a) $\epsilon =-1\times 10^{-4}$
$\delta=1\times 10^{-4}$ for $O2$ and $\delta=0$ for $O1$ orbit and
(b) $\epsilon =1\times 10^{-4}$,
$\delta=5\times 10^{-5}$ for $O2$ and $\delta=1\times 10^{-6}$ for $O1$ orbit;
we first calculate  period-5 orbit ($\bar X_n$) from eqn.(31) and 
then take the initial condition of the trajectory as $X_n=\bar X_n +\delta$ 
($\delta$ being taken nonzero only for $n=4$).

\bigskip

\noindent {\bf Fig.17} ~~2-dim projection of two trajectories  starting from
close to the period-5 orbits ($O1$ and $O2$) in type II bifurcation near the
$secondary$ bifurcation  with $\beta=-2$, $\alpha =1$ and 
$\epsilon =-5\times 10^{-3}$ corresponding to (a) $q =6\times 10^{-2}$
$\delta=5\times 10^{-5}$ for $O2$ and $\delta=1\times 10^{-4}$ for $O1$ orbit 
and (b) $q =4.806\times 10^{-2}$,
$\delta=0$ for $O2$ and $O1$ orbit~;
we first calculate  period-5 orbit ($\bar X_n$) from eqn.(31) and 
then take the initial condition of the trajectory as $X_n=\bar X_n +\delta$ 
($\delta$ being taken nonzero only for $n=4$).

\bigskip

\noindent {\bf Fig. 18} ~~2-dim projection of two trajectories  starting from
close to the period-6 orbits ($O1$ and $O2$) in type I bifurcation near the
bifurcation point with $\beta=1$,
$\alpha =1$ and $q=-2\times 10^{-1}$ corresponding to (a) $\epsilon =-1\times 10^{-2}$
$\delta=1\times 10^{-4}$ for $O1$ and $\delta=0$ for $O2$ orbit and
(b) $\epsilon =1\times 10^{-3}$,
$\delta=2\times 10^{-4}$ for $O1$ and $\delta=0$ for $O2$ orbit;
we first calculate  period-5 orbit ($\bar X_n$) from eqn.(41) and 
then take the initial condition of the trajectory as 
$X_n=\bar X_n +2n(-1)^n\delta cos(2n\phi)$
($\phi$=0 for $O1$ and 
${\pi\over 6}$ for $O2$). 

\bigskip

\noindent {\bf Fig. 19} ~~2-dim projection of two trajectories  starting from
close to the period-6 orbits ($O1$ and $O2$) in type II bifurcation near the
$secondary$ bifurcation  with $\beta=-4$, $\alpha =1$ and 
$\epsilon =-1\times 10^{-2}$ corresponding to (a) $q =-6.8\times 10^{-2}$
$\delta=-1.6\times 10^{-6}$ for $O2$ and $\delta=1\times 10^{-6}$ for $O1$ orbit 
and (b) $q =-6.2\times 10^{-2}$,
$\delta=0$ for $O2$ and $O1$ orbit~;
we first calculate  period-6 orbit ($\bar X_n$) from eqn.(41) and 
then take the initial condition of the trajectory as 
$X_n=\bar X_n +2n(-1)^n\delta cos(2n\phi)$. ($\phi$=0 for $O1$ and 
${\pi\over 6}$ for $O2$). The inset shows one of the stable period-6
orbit surrounded by the sepratrix of the unstable period-6 point.

\bigskip

\noindent {\bf Fig. 20} ~~(a)~2-dim projection of a trajectory around one of
the period-4 points of the $O1$ orbit on the subthreshold side of the secondary
bifurcation, with
$\epsilon =-10^{-5}$, $\beta=-1$, $\gamma=-3.02$ and
$q=-2.585\times 10^{-3}$~;
we first calculate an initial condition ($X_1, X_2, X_3, X_4$) 
corresponding to one member belonging to a family of invariant curves, with
$\hat\phi =1.18\times 10^{-4}$ (see ref.s [11,15] for explanation~); a slightly
different initial condition is then taken with $X_1, X_2, X_3$ unchanged
and $X_4$ replaced by $X_4 +\delta$ ($\delta =3\times 10^{-10}$) and
iterations performed; the trajectory winds on a 2-torus around the invariant
curve.  (b) A  2-torus on the superthreshold side near a period-4
point of the $O1$ orbit~; with
$\epsilon =-10^{-5}$, $\beta=-1$, $\gamma=-10^{-2}$ and
$q=-2.577\times 10^{-3}$,
$\hat\phi =10^{-5}$,
$\delta=10^{-9}$ (see legend for (a) ).
\end